\documentclass[preprintnumbers,amsmath,amssymb]{revtex4}
\usepackage{graphicx}

\usepackage{dcolumn}
\usepackage{bm}

\usepackage{graphicx}
\usepackage{epsfig}

\usepackage{amsmath,amsthm}
\usepackage{amssymb,latexsym}

\usepackage{setspace}
\usepackage[mathscr]{eucal}
\usepackage{amsthm}
\usepackage{eepic}
\usepackage{psfrag}

\numberwithin{equation}{section}

\newcommand{\bee}{\begin{equation}}
\newcommand{\ene}{\end{equation}}

\begin{document}

\title{Numerical Simulations of Snake Dissipative Solitons in Complex Cubic-Quintic Ginzburg-Landau Equation}

\vskip0.4in

\author{Stefan C. Mancas\\
and\\
Harihar Khanal \\
Embry-Riddle Aeronautical University, Daytona Beach, USA \\
mancass@erau.edu\\
khana66a@erau.edu}

\begin{abstract}

\vspace{.1in}

Numerical simulations of the complex cubic-quintic Ginzburg-Landau equation (CCQGLE),
a canonical equation governing the weakly nonlinear behavior of dissipative systems in a wide variety of disciplines, reveal five entirely novel classes of pulse or solitary waves solutions, viz. pulsating, creeping, snaking, erupting, and chaotical solitons \cite{KM1}.
Here, we develop a theoretical framework for analyzing the full spatio-temporal structure of
one class of dissipative solution (snaking soliton) of the CCQGLE using the variational approximation technique and the dynamical systems theory. The qualitative behavior of the snaking soliton is investigated using the numerical simulations of (a) the full nonlinear complex partial differential equation and (b) a system of three ordinary differential equations resulting from the variational approximation.

\end{abstract}

\maketitle
\section{Introduction}\label{S:1}
The complex cubic-quintic Ginzburg-Landau equation (CCQGLE)
is one of the most widely studied nonlinear equations. In
fluid mechanics, it is also often referred to as the
Newell-Whitehead equation after the authors who derived it in the
context of B\'enard convection. Many basic properties of the
equation and its solutions are reviewed in \cite{Aranson},
together with applications to a vast variety of phenomena including
nonlinear waves, second-order phase transitions, superconductivity,
superfluidity, Bose-Einstein condensation, liquid crystals and
string theory.

Due to the highly nonlinear nature of the CCQGLE, analytic solutions
have been limited to special cases. Thus, its
numerical approximations have become valuable tools in order to
better understand the theoretical insight into the problem.
Numerical simulations in \cite{KM1} identify five entirely novel classes of pulse or
solitary waves solutions, viz. pulsating, creeping, snaking, erupting,
and chaotical solitons of the CCQGLE.
In contrast to the regular solitary waves investigated in numerous
integrable and non-integrable systems over the last three decades,
these dissipative solitons are not stationary in time. Rather, they
are spatially confined pulse-type structures whose envelopes
exhibit complicated temporal dynamics.

In this paper we describe a theoretical framework for
analyzing the full spatiotemporal structure of one class of solitary
waves (snaking soliton) of the CCQGLE, and present numerical simulations
resulting from  the full nonlinear partial differential equation, as well as systems of ODEs from the variational approximation technique.

First, we develop and discuss a variational formalism within which to explore the various classes of dissipative solitons. We find a suitable Lagrangian based on ansatz. Then the resulting Euler-Lagrange equations are treated in a completely novel way. Rather than considering on the stable fixed points which correspond to the well-known stationary solitons or plain pulses, we use dynamical systems theory to focus on more complex attractors viz. periodic, quasiperiodic, and chaotic ones. Periodic evolution of the trial function parameters on a stable periodic attractor would yield solitons whose amplitudes are non-stationary or time dependent. In particular, pulsating, snaking (and less easily, creeping) dissipative solitons may be treated using stable periodic attractors of various trial function parameters. Chaotic evolution of the trial function parameters would yield to chaotic solitary waves.
In the language of the Los Alamos school, the fully spatiotemporal approach followed here may be said to be the ``collective coordinates" formulation. In other words, we consider a pulse or solitary wave at any time as a coherent collective entity (or coordinate). This solitary wave is then temporally modulated. The spatial approach proposed, and explored, in this paper is the variational method. However, the method is very significantly and non-trivially generalized from all earlier applications to deal with our novel classes of dissipative solitary waves.

Next, we solve (a) the initial boundary value problem of the CCQGLE and (b) the initial value problem of the Euler-Lagrange system resulting from the variational approximation numerically.
The numerical scheme for the PDE is based on a Finite Difference discretization of the partial differential equation
with a low storage varient explicit Runge-Kutta third order method. The Euler-Lagrange system of ODEs is solved numerically using the MATLAB's built-in function \verb+ode45+.

The remainder of this paper is organized as follows. Section \S2
describes briefly the CCQGLE as well as the classes of its soliton solutions. An ODE
system resulting from the generalized variational formulation of the
CCQGLE and Hopf bifurcation theory are also presented here. In section \S3 numerical
methods for the IVP and BVP are discussed. Sections \S4 and \S5 deal with the simulation results from previous section using the PDE and the variational formulation. Finally, in section \S5 we summarize the results and conclude the paper outlining some further research directions.
\section {Description of CCQGLE}
An important element  in the long time dynamics of pattern forming
systems is a class of solutions which we call ``coherent
structures" or solitons. These structure could be a profile of
light intensity, temperature, magnetic field, etc. A dissipative
soliton is localized and exists for an extended period of time. Its
parts are experiencing gain/loss of energy or mass with the medium.
Thus, energy and matter can flow into the system through its
boundaries. As long as the parameters in the system stay constant,
the structure that could evolve by changing shape exists
indefinitely in time. The structure disappears when the source is
switched off, or if the parameters are moved outside the range of
existence of the soliton solutions. Since these dissipative systems
include energy exchange with external sources, they are no longer
Hamiltonian, and the solitons solutions in these systems are also
qualitatively different from those in Hamiltonian systems. In
Hamiltonian systems, soliton solutions appear as a result of balance
between diffraction (dispersion) and nonlinearity. Diffraction
spreads a beam while nonlinearity will focus it and make it
narrower. The balance between the two results in stationary solitary
wave solutions, which usually form a one parameter family. In
dissipative systems with gain and loss, in order to have stationary
solutions, the gain and loss must be also balanced. This additional
balance results in solutions which are fixed. Then the shape,
amplitude and the width are all completely fixed by the parameters
of the dissipative equation. However, the solitons, when they exist, can
again be considered as ``modes" of dissipative systems just as for
non-dissipative ones.

Recent perturbative treatments based on expressions about the
nonlinear Schr\"odinger equations are generalized to perturbations
of the cubic-quintic and derivative Schr\"odinger equations. The
cubic Ginzburg-Landau admits a selected range of exact soliton
solutions \cite{Maruno}.
These exist when certain relations
between parameters are satisfied. However, this certainly does not
imply that the equations are integrable. Dissipative systems, as it
is the case with Burger's equation are integrable. In reality,
general dissipative nonlinear PDEs cannot be reduced to linear equations in
any known way, therefore an insight to the type of solutions that
the equation has may be based on numerical simulations.

The simplest mode that can account for this type of behavior is
the Ginzburg-Landau model, with the corresponding cubic-quintic
equations in the form \cite{Saarloos}
\begin{equation}\label{2.1}
\partial_tA=\epsilon A+(b_1+ic_1)\partial_x^2A-(b_3-ic_3) |A|^2A-(b_5-ic_5) |A|^4 A.
\end{equation}
The present study will confine itself to spatially infinite systems
in one dimension  and will focus primarily on a spatio-temporal
behavior of dissipative solitons. One way to approach this problem
and to arrive at \eqref{2.1} is to start with the well known
modified Schr\"odinger equation
\begin{equation}\label{2.2}
\partial_tA=i c_1 \partial_x^2A+i c_3  |A|^2A +i c_5  |A|^4A +\partial_x[(s_0+s_2|A|^2)A],
\end{equation}
and to perturb with a dissipative term \cite{Fauve} in  special
parameter regimes, with terms on the rhs. of \eqref{2.2} of the form
$f(A)=\epsilon A+b_1\partial_x^2A-b_3 |A|^2A-b_5|A|^4A$.

The unperturbed system \eqref{2.2} leads to an integrable dynamical
system whose orbits can be calculated analytically \cite{Cariello}.

The interpretation of the system's parameters in \eqref{2.1} depends
on the particular field of work. For example, in optics, $t$ is the
propagation distance or the cavity number, $x$ is the transverse
variable, the angular spectral gain or loss is identified by $b_1$,
$c_1$ is the second-order diffraction or linear dispersion, $b_3$
is the nonlinear gain or 2-photon absorption if negative, $c_3$ is
the nonlinear dispersion, $\epsilon$ is the difference between
linear gain and loss, $b_5$ represents the saturation of the
nonlinear gain, and $c_5$ the saturation of the nonlinear refractive
index. In physical problems, the quintic nonlinearity is even of
higher importance than the cubic one, as it is responsible for
stability of localized solutions.

A special case of \eqref{2.1} is the nonlinear Schr\"odinger equation
($\epsilon=b_1=b_3=b_5=c_5=0$, $c_1=1$)
\begin{equation}\label{NLS}
\partial_tA=i  \partial_x^2A+i c_3 |A|^2A
\end{equation}
which is both Hamiltonian and integrable. Its extension, the
quintic-cubic Scr\"odinger equation ($\epsilon=b_1=b_3=b_5=0$, $c_1=1$)
\begin{equation}\label{QCS}
\partial_tA=i  \partial_x^2A+i c_3 |A|^2A + i c_5 |A|^4A
\end{equation}
is Hamiltonian but non-integrable. Other interesting cases are the
derivative nonlinear Schr\"odinger equation
\begin{equation}\label{DNS}
\partial_tA=i  \partial_x^2A+s_0 \partial_x A+s_2\partial_x(|A|^2A)
\end{equation}
and the combinations of last two, which is the quintic derivative
Scr\"odinger equation
\begin{equation}\label{QDS}
\partial_tA=i  \partial_x^2A+s_0 \partial_x A+s_2\partial_x(|A|^2A)+ic_3 |A|^2A+ ic_5|A|^4A.
\end{equation}
\subsection{Euler-Lagrange Equations}
Employing the generalized variational formulations and proceeding as in
\cite{Kaup:1}, the Lagrangian for the
CCQGLE \eqref{2.1} may be written as \cite{Mancas:3}
\begin{align}\label{3.1}
\mathscr{L}&=r^*\big[\partial_tA-\epsilon A-(b_1+ic_1)\partial_x^2A+(b_3-ic_3) |A|^2A+(b_5-ic_5) |A|^4 A\big]\notag\\
&+r\big[\partial_tA^*-\epsilon
A^*-(b_1-ic_1)\partial_x^2A^*+(b_3+ic_3) |A|^2A^*+(b_5+ic_5) |A|^4
A^*\big].
\end{align}
Here $r$ is the usual auxiliary equation employed in \cite{Kaup:1}
and it satisfies a perturbative evolution equation dual to the CCQGLE
with all non-Hamiltonian terms reversed in sign.
Choosing the single-humped trial functions of the form:
\begin{eqnarray}
A(x,t) &=& A_1(t)e^{-\sigma_1(t)^2\lbrack x-\phi_1(t) \rbrack^2}e^{i\alpha_1(t)}e^{i \psi(t)^2 x^2}\label{3.2}\\
r(x,t) &=& e^{-\sigma_2(t)^2\lbrack x-\phi_2(t) \rbrack^2}e^{i\alpha_2(t)}\label{3.3}
\end{eqnarray}
where the $A_1(t)$ is the amplitude, the $\sigma_i(t)$'s are the
inverse widths, $\phi_i(t)$'s are the positions (with $\dot{\phi}_i(t)$ the speed), $e^{i \psi(t)^2
x^2}$ represents the chirp, $\alpha_i(t)$'s are the phases of the
solitons and are all allowed to vary arbitrarily in time.
For now, the chirp terms are omitted for simplicity. Substituting \eqref{3.2}/\eqref{3.3} in \eqref{3.1} the effective or averaged Lagrangian is
\begin{eqnarray}\label{3.4}
L_{EFF}&=&\int_{-\infty}^{\infty}\mathscr{L}dx=2\sqrt{\pi}\Bigg\{-\frac{e^{-\frac{\sigma_1(t)^2\sigma_2(t)^2[\phi_1(t)-\phi_2(t)]^2}{\sigma_1(t)^2+\sigma_2(t)^2}}}{[\sigma_1(t)^2+\sigma_2(t)^2]^\frac12}\epsilon A_1(t)\cos[\alpha_1(t)-\alpha_2(t)] \nonumber \\
&+&\frac{e^{-\frac{3\sigma_1(t)^2\sigma_2(t)^2[\phi_1(t)-\phi_2(t)]^2}{3\sigma_1(t)^2+\sigma_2(t)^2}} }{\Big[3\sigma_1(t)^2+\sigma_2(t)^2\Big]^\frac12}A_1(t)^3\Bigg[b_3\cos[\alpha_1(t)-\alpha_2(t)]+c_3\sin[\alpha_1(t)-\alpha_2(t)]\Bigg] \nonumber \\
&+&\frac{e^{-\frac{5\sigma_1(t)^2\sigma_2(t)^2[\phi_1(t)-\phi_2(t)]^2}{5\sigma_1(t)^2+\sigma_2(t)^2}} }{\Big[5\sigma_1(t)^2+\sigma_2(t)^2\Big]^\frac12}A_1(t)^5\Bigg[b_5\cos[\alpha_1(t)-\alpha_2(t)]+c_5\sin[\alpha_1(t)-\alpha_2(t)]\Bigg] \nonumber\\
&+&\frac{e^{-\frac{\sigma_1(t)^2\sigma_2(t)^2[\phi_1(t)-\phi_2(t)]^2}{\sigma_1(t)^2+\sigma_2(t)^2}}}{\Big[\sigma_1(t)^2+\sigma_2(t)^2\Big]^\frac52}\Bigg[\cos[\alpha_1(t)-\alpha_2(t)][\sigma_1(t)^2+\sigma_2(t)^2]^2\dot{A}_1(t)\nonumber\\
&+&A_1(t)\Bigg(-2\sigma_1(t)^2\sigma_2(t)^2\Big[b_1\cos[\alpha_1(t)-\alpha_2(t)]-c_1\sin[\alpha_1(t)-\alpha_2(t)]\Big]\Big[-\sigma_2(t)^2 \nonumber \\
&+&\sigma_1(t)^2[-1+2\sigma_2(t)^2[\phi_1(t)-\phi_2(t)]^2]\Big]-\dot{\alpha_1}(t)\sin[\alpha_1(t)-\alpha_2(t)][\sigma_1(t)^2+\sigma_2(t)^2]^2\nonumber \\
&-&\sigma_1(t)\dot{\sigma_1}(t)\cos[\alpha_1(t)-\alpha_2(t)]\Big[\sigma_1(t)^2+\sigma_2(t)^2+2\sigma_2(t)^4[\phi_1(t)-\phi_2(t)]^2\Big] \nonumber \\
&-&2\dot{\phi_1}(t)\sigma_1(t)^2\sigma_2(t)^2[\phi_1(t)-\phi_2(t)][\sigma_1(t)^2+\sigma_2(t)^2]\cos[\alpha_1(t)-\alpha_2(t)]\Bigg)\Bigg]\Bigg\}.
\end{eqnarray}
Since \eqref{3.4} reveals that only the relative phase
$ \alpha(t)=\alpha_1(t)-\alpha_2(t)$, and relative velocity $ \phi(t)=\phi_1(t)-\phi_2(t)$ of $A(x,t)$ and $r(x,t)$ are relevant, we henceforth rescale them according to
\begin{equation}\label{assumption}
\alpha_1(t)=\alpha(t), \alpha_2(t)=0;\, \phi_1(t)=\phi(t), \phi_2(t)=0.
\end{equation} Also, for algebraic tractability, we have found it necessary to assume
\begin{equation}\label{3.6}
\sigma_2(t)=\sigma_1(t)\equiv \sigma(t).
\end{equation} While this ties the widths of the $A(x,t)$ and $r(x,t)$ fields together, the loss of generality is acceptable since the field $r(x,t)$ has no real physical significance.
Moreover, for the snaking soliton solutions, we choose $\sigma(t)=2/\phi(t)$. By using all these assumptions, the trial functions \eqref{3.2}-\eqref{3.3} become
\begin{eqnarray}
A(x,t) &=& A_1(t)e^{-\frac{4}{\phi(t)^2}[x-\phi(t)]^2}e^{i\alpha(t)} \label{5.2} \\
r(x,t) &=& e^{-\frac{4}{\phi(t)^2}x^2}.\label{5.3}
\end{eqnarray}
Substituting the last two equations into \eqref{3.4}, the effective Lagrangian becomes
\begin{eqnarray}\label{5.4}
L_{EFF} & = & \frac{\sqrt{\pi}}{12 e ^{\frac{10}{3}}\phi(t)}\Bigg[6 e^\frac13 A_1(t)^3 \phi(t)^2 \big(b_3 \cos \alpha(t)+c_3 \sin \alpha(t) \big) \nonumber \\
&+& 2 \sqrt 6  A_1(t)^5\phi(t)^2 \big( b_5 \cos \alpha(t)+c_5 \sin \alpha(t) \big)\nonumber \\
&-& 3 \sqrt 2e^{-\frac43}\bigg(-2 A_1(t)\sin \alpha(t)\big(-12 c_1+\phi(t)^2\dot{\alpha(t)}\big)\nonumber \\
&+& \cos \alpha(t) \big(-2 \phi^2(t)\dot{\alpha}(t)+A_1(t)(24 b_1+2 \epsilon\phi^2(t)+3\phi(t)\dot{\phi}(t)) \big)\bigg) \Bigg].
\end{eqnarray}
We are left with three parameters $A_1(t)$,  $\phi(t)$ and $\alpha(t)$ in $L_{EFF}$.
Varying these parameters by using calculus of variations, we obtain
\begin{equation}\label{4.5.1}
\frac{\partial L_{EFF}}{\partial \star(t)}-\frac{\mathrm{d}}{\mathrm{dt}}\Big(\frac{\partial L_{EFF}}{\partial \dot{\star}(t)}\Big)=0,
\end{equation} where $\star$ refers to
$A_1$, $\alpha$, or $\phi$.
Solving for $\dot{\star}(t)$, we obtain the following system of three ordinary differential equations,
\begin{align}
\dot{A_1}(t)&=f_1 \lbrack A_1(t),\alpha(t),\phi(t) \rbrack \notag\\
\dot{\alpha}(t)&=f_2  \lbrack A_1(t),\alpha(t),\phi(t) \rbrack \notag\\
\dot{\phi}(t)&=f_3  \lbrack A_1(t),\alpha(t),\phi(t) \rbrack
\label{4.5}
\end{align}
where the right hand side $f_1, f_2, f_3$ of \eqref{4.5} are given by
\begin{eqnarray}
f_1 \,\,=\!\!\!\! &&
\frac15A_1(t)\sec\alpha(t)(3456c_1e^{8/3}\sin\alpha(t)(-b_1\cos\alpha(t)+c_1\sin\alpha(t)) \nonumber \\
&&
+A_1(t)^2\phi(t)^2(2\sqrt{6}e^{1/3}A_1(t)^4(5b_3b_5
+17c_3c_5+(3b_3b_5-19c_3c_5)\cos(2\alpha(t))\nonumber \\
&&+(-2b_3b_5+2c_3c_5)\cos(4\alpha(t))
+(b_5c_3+b_3c_5)(11\sin(2\alpha(t))-2\sin(4\alpha(t))))\phi(t)^2
\nonumber \\
& &
+4A_1(t)^6(4b_5^2+13c_5^2+(2b_5^2-15c_5^2)\cos(2\alpha(t))
+2(-b_5^2+c_5^2)\cos(4\alpha(t))\nonumber \\
&&+17b_5c_5\sin(2\alpha(t)) -4b_5c_5\sin(4\alpha(t)))\phi(t)^2+3e^{2/3}A_1(t)^2(3b_3^2 +11c_3^2
\nonumber \\
&&
+2(b_3^2-6c_3^2)\cos(2\alpha(t))+(-b_3^2+c_3^2)\cos(4\alpha(t))
-2b_3c_3(-7\sin(2\alpha(t))\nonumber \\
&&+\sin(4\alpha(t))))\phi(t)^2
+6\sqrt{2}e^{5/3}(6b_1b_3-18c_1c_3+6(b_1b_3+c_1c_3)(2\cos(2\alpha(t))
\nonumber \\
&&
+\cos(4\alpha(t)))-48(b_3c_1-b_1c_3)\cos\alpha(t)^3\sin\alpha(t)
-2\epsilon\cos\alpha(t)(b_3\cos\alpha(t)\nonumber \\
&&+3c_3\sin\alpha(t))\phi(t)^2)+4\sqrt{3}e^{4/3}A_1(t)^2(12b_1b_5-36c_1c_5
\nonumber \\
&&
+12(b_1b_5+c_1c_5)(2\cos(2\alpha(t))+\cos(4\alpha(t))) -96(b_5c_1-b_1c_5)\cos\alpha(t)^3\sin\alpha(t)
\nonumber \\
&&
-2\epsilon\cos\alpha(t)(2b_5\cos\alpha(t)
+5c_5\sin\alpha(t))\phi(t)^2)))e^{-4/3}/\phi(t)^2/(144c_1e^{4/3}\sin\alpha(t)
\nonumber \\
&&
+A_1(t)^2(-3\sqrt{2}e^{1/3}(-3b_3\cos\alpha(t)+b_3\cos(3\alpha(t))
-11c_3\sin\alpha(t)+c_3\sin(3\alpha(t)))
\nonumber \\
&&+4\sqrt{3}A_1(t)^2(4b_5\cos\alpha(t)-2b_5\cos(3\alpha(t))
+13c_5\sin\alpha(t)\nonumber \\
&&-2c_5\sin(3\alpha(t))))\phi(t)^2)
\\ \nonumber
\end{eqnarray}
\begin{eqnarray}
f_2 \,\,= \!\!\!\! &&
\frac{1}{15}(-72e^{4/3}(b_1-c_1\tan\alpha(t))
+(-6e^{4/3}\epsilon+9\sqrt{2}e^{1/3}A_1(t)^2(b_3
+c_3\tan\alpha(t))\nonumber \\
&&+10\sqrt{3}A_1(t)^4(b_5+c_5\tan\alpha(t)))\phi(t)^2)e^{-4/3}/\phi(t)
\\ \nonumber
\end{eqnarray}
\begin{eqnarray}
f_3 \,\,= \!\!\!\! &&
\frac{2}{5}(-144c_1e^{7/3}\epsilon\cos\alpha(t)
-\sqrt{6}A_1(t)^6((b_5c_3-3b_3c_5)\cos\alpha(t)
\nonumber \\
&&
+(b_5c_3+b_3c_5)\cos(3\alpha(t))-2(b_3b_5+c_3c_5 +(b_3b_5-c_3c_5)\cos(2\alpha(t)))\sin\alpha(t))\phi(t)^2
\nonumber \\
&&
+6\sqrt{2}e^{4/3}A_1(t)^2(12(b_3c_1+3b_1c_3)\cos\alpha(t)
+12(b_3c_1-b_1c_3)\cos(3\alpha(t))
\nonumber \\
&&+12(b_1b_3-3c_1c_3)\sin\alpha(t)+12(b_1b_3+c_1c_3)\sin(3\alpha(t))+\epsilon\cos\alpha(t)(c_3(-2\nonumber \\
&&+\cos(2\alpha(t)))-b_3\sin(2\alpha(t)))\phi(t)^2)+8\sqrt{3}e A_1(t)^4(36b_1c_5\cos\alpha(t)
\nonumber \\
&&
+18(b_5c_1-b_1c_5)\cos(3\alpha(t))+18(b_1b_5-3c_1c_5)\sin\alpha(t)
+18(b_1b_5+c_1c_5)\sin(3\alpha(t))\nonumber \\
&&
+\epsilon\cos\alpha(t)(-3c_5+2c_5\cos(2\alpha(t))
-2b_5\sin(2\alpha(t)))\phi(t)^2))e^{-1}/(144c_1e^{4/3}\sin\alpha(t)
\nonumber \\
&&
+A_1(t)^2(-4\sqrt{2}e^{1/3}(-3b_3\cos\alpha(t)+b_3\cos(3\alpha(t))
-11c_3\sin\alpha(t)
\nonumber \\
&&+c_3\sin(3\alpha(t)))+4\sqrt{3}A_1(t)^2(4b_5\cos\alpha(t)
-2b_5\cos(3\alpha(t))+13c_5\sin\alpha(t)
\nonumber \\
&&-2c_5\sin(3\alpha(t))))\phi(t)^2)
\end{eqnarray}
\subsection{Hopf Bifurcations}
The Euler-Lagrange equations  \eqref{4.5} are treated in a completely novel way. Rather than considering the stable fixed points which correspond to the well-known stationary solitons or plain pulses, we use Hopf bifurcation theory to focus on periodic attractors. Periodic evolution of the trial function parameters on stable periodic attractors yields the pulsating soliton whose amplitude is non-stationary or time dependent.

We derive the conditions for the temporal Hopf bifurcations of the fixed points. The conditions for supercritical temporal Hopf bifurcations, leading to stable periodic orbits of $A_1(t)$, $\phi(t)$, and $\alpha(t)$ can be evaluated using the method of Multiple Scales, as in \cite{Mancas:4}. These are the conditions or parameter regimes which exhibit stable periodic oscillations, and hence stable pulsating solitons will exist within our variational formulation. Note that, it is easy to verify numerically, periodic oscillations of $A_1(t)$, $\phi(t)$, and $\alpha(t)$, correspond to a spatiotemporal pulsating soliton structure of the $|A(x,t)|$ given by \eqref{3.2}.

The fixed points of \eqref{4.5} are given by a complicated system of transcendental equations. These are solved numerically to obtain results for each particular case.

For a typical fixed point, the characteristic polynomial of the Jacobian matrix of a fixed point of \eqref{4.5} may be expressed as
\begin{equation} \label{4.6}
\lambda^3+\delta_1\lambda^2+\delta_2\lambda +\delta_3=0
\end{equation}
where
$\delta_i$ with $i=1...3$ depend on the system parameters and the fixed points. Since these are extremely involved, we omit the actual expressions, and evaluate them numerically where needed.

To be a stable fixed point within the linearized analysis, all the eigenvalues must have negative real parts. Using the Routh-Hurwitz criterion, the necessary and sufficient conditions for \eqref{4.6} to have $Re(\lambda_{1,2,3})<0$ are:
\begin{equation}\label{4.7}
\delta_1>0,\quad\delta_3>0,\quad\delta_1\delta_2-\delta_3>0.
\end{equation}

On the contrary, one may have the onset of instability of the plane wave solution occurring in one of the 
two ways. In the first, one root of \eqref{4.5} (or one eigenvalue of the Jacobian) becomes non-hyperbolic 
by going through zero for
\begin{equation} \label{4.8}
\delta_3=0.
\end{equation}
Equation \eqref{4.8} is thus the condition for the onset of ``static" instability of the plane wave. Whether this bifurcation is a pitchfork or transcritical one, and its subcritical or supercritical nature, may be readily determined by deriving an appropriate canonical system in the vicinity of \eqref{4.8} using any of a variety of normal form or perturbation methods.

One may also have the onset of dynamic instability (``flutter" in the language of Applied Mechanics) when a pair of eigenvalues of the Jacobian become purely imaginary. The consequent Hopf bifurcation at
\begin{equation}\label{4.9}
\delta_1\delta_2-\delta_3=0
\end{equation}
leads to the onset of periodic solutions of \eqref{4.5} (dynamic instability or ''flutter").
\section{Numerical Methods}
\subsection{The Initial Boundary Value Problem of the CCQGLE}
To obtain the snaking soliton solutions that
are periodic and exponentially decaying at infinity,
we develop numerical schemes for the IBVP 
of (\ref{2.1}) with $A(a,\,\cdot)=A(b,\,\cdot)$,
and $A(0,\,\cdot)=A_0(\cdot)$
based on the finite difference method as described below.

Let $h=(b-a)/n$ be the grid spacing and $x_j=a+jh, \, j=0,1,\cdots,n$
be the grid points.
Define $A_j(t)$ as an approximation to $A(x_j,t),\,j=0,1,\cdots,n$,
${\bf A}(t)= (A_1(t),\cdots,A_{n-1}(t))^T$
and ${\bf A}_0(x)= (A_0(x_1),\cdots,A_{0}(x_{n-1}))^T$.
Let $F(A_j(t))=
\epsilon A_j(t) -(b_3-ic_3)|A_j(t)|^2A_j(t)-(b_5-ic_5)|A_j(t)|^4A_j(t)$.
Using the central difference approximation to $\partial_x^2 A(x_j,t)$,
we write the semi-discretization of the IBVP of (\ref{2.1}) as
the following system of ODEs
\begin{eqnarray}
&\dot{\bf A}(t) = {\cal B}{\bf A}(t) + {\bf F}({\bf A}(t)), \, \forall t>0,
& \label{eq:4}\\
&{\bf A}(0)={\bf A}_0, & \label{eq:5}
\end{eqnarray}
where ${\bf F}({\bf A}(t))=\left(F(A_1(t)),\,\cdots,\,F(A_n(t))\right)^T$.
The upper dot in (\ref{eq:4}) indicates derivative with respect to $t$ and
${\cal B}$ is the finite difference matrix.
To construct an integration scheme to solve the ODE system
(\ref{eq:4})-(\ref{eq:5}), let  $t_{n+1} = t_n +\Delta t$,
and let ${\bf A}^{n}$ denote the value of the
variable ${\bf A}$ at time $t_n$.
Employing a low storage varient third-order Runge-Kutta scheme
\cite{Williamson}, we write the fully discrete system as
\begin{equation}\label{RK3}
\begin{array}{lll}
&{\bf Q}_1 = \Delta t {\bf G}({\bf A}^n), &
{\bf A}_1 = {\bf A}^n + \frac13 {\bf Q}_1, \\
&{\bf Q}_2 = \Delta t {\bf G}({\bf A}_1)- \frac59{\bf Q}_1, &
  {\bf A}_2 = {\bf A}_1 + \frac{15}{16} {\bf Q}_2,\\ 
&{\bf Q}_3 = \Delta t {\bf G}({\bf A}_2)-\frac{153}{128}{\bf Q}_2, &
{\bf A}^{n+1} = {\bf A}_2 + \frac{8}{15} {\bf Q}_3, \\ 
\end{array}
\end{equation}
where ${\bf G}({\bf A}^n)={\cal B}{\bf A}^n + {\bf F}({\bf A}^n)$.
The numerical code 
is parallelized for distributed memory clusters of
processors or heterogeneous networked computers using the
MPI (Message Passing Interface) library and implemented in FORTRAN.

Computations were performed on a Linux cluster
(zeus.db.erau.edu: 256 Intel Xeon 3.2GHz 1024 KB cache 4GB with Myrinet MX,
GNU Linux) at Embry-Riddle Aeronautical University.
\subsection{The Initial Value Problem of Euler-Lagrange System}
The system of ODEs (\ref{4.5}) resulting from the variational approximation
to the PDE is solved numerically using the MATLAB's built-in function \verb+ode45+
(adaptive fourth- and fifth-order Runge-Kutta-Fehlberg method).
We use equilibrium solution
$A_1(0)=A_1^0, \, \alpha(0)=\alpha^0, \, \phi(0)=\phi^0$ of the system (\ref{4.5})
as the initial conditions.
The numerical solutions $A_1^n$, $\alpha^n$, $\phi^n$ giving the
approximations to $A_1(t_n)$, $\alpha(t_n)$, $\phi(t_n)$ are used to
evaluate $A(x_j,t_n)$, approximation to the snaking soliton solution to
the CCQGLE, in the spatial grids $x_j$ at time $t_n$
using the following formula
\begin{equation*}
A(x_j,t_n)=
A_1^n e^{-\sigma^2\left(x_j-\phi^n\right)^2}e^{i\alpha^n},
\end{equation*}
where $\sigma=2/\phi^n$.
\section{Results and Discussion}
\subsection{Simulations of Snake Solitons using the PDE}
The numerical scheme for the PDE described in section \S3
is implemented with the following initial amplitude profile
\begin{equation}\label{init}
A(x,0) = A_1 e^{-\sigma^2(x-\phi)^2}e^{i\alpha}.
\end{equation}
We use $A_1=0.583236$, $\phi=1.05969$, $\alpha=0.185515$ and $\sigma=1.8873$ 
as the typical values of the parameters in (\ref{init}). These are the values of the fixed point as we will see in the next section. For the system parameters of the PDE (\ref{2.1}), initially we use the values listed in Table 1 (see the row for snaking soliton). 
In \cite{KM1} five novel classes of dissipative soliton solutions
viz. pulsating, creeping, snaking, exploding and chaotical were obtained by
numerical simulations of the CCQGLE (\ref{2.1}).
Here we present a theoretical formulation to one class of the solutions, the snake, we perform independent simulations on the full PDE from the ones shown previously in \cite{Akhmediev:1}, and we also compare the result with the simulations from the variational approximation.
First, for the simulations on the PDE, we fix a set of system parameters $\epsilon$, $b_1$, $b_3$, $b_5$, $c_1$, $c_3$, $c_5$
for the snake soliton from the Table 1. Then, we study the qualitative behavior
of the snaking soliton by varying one parameter at a time.
Then,  we  present the spatio-temporal structure of the snaking soliton
in detail using the numerical simulations of the variational approximation to CCQGLE in the next section.
\begin{figure}[ht]\label{figure1}
\begin{center}
\resizebox*{0.3\textheight}{!}{\rotatebox{0}
{\includegraphics{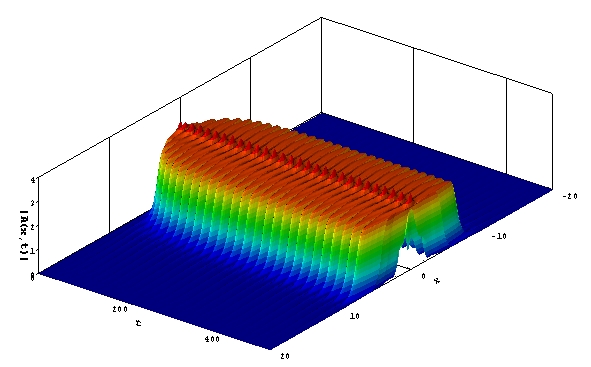}}}
\end{center}
\caption{Numerical simulation of snaking soliton using the PDE: unperturbed case}
\end{figure}
\begin{table}[ht]
\begin{center}
\caption{Parameters of the CCQGLE}
\renewcommand{\baselinestretch}{1.0}
\begin{tabular}{|l|ccccccc|}
\hline
Solitons & $\epsilon$ & $b_1$ & $c_1$ & $b_3$ & $c_3$ & $b_5$ & $c_5$ \\
\hline
\hline
Pulsating & -0.1 & 0.080 & 0.5 & -0.660 & 1 & 0.10 & -0.100 \\ 
\hline
Creeping & -0.1 & 0.101 & 0.5 & -1.300 & 1 & 0.30 & -0.101 \\ 
\hline
Snaking 
& -0.1 & 0.080 & 0.5 & -0.835 & 1 & 0.11 & -0.080  \\ 
\hline
\hline
Exploding & -0.1 & 0.125 & 0.5 & -1.000 & 1 & 0.10 & -0.600 \\ 
\hline
Chaotical & -0.1 & 0.125 & 0.5 & -0.300 & 1 & 0.10 & -1.000  \\
\hline
\end{tabular}
\renewcommand{\baselinestretch}{1.5}
\end{center}
\vspace*{-.1in}
\end{table}
The spatiotemporal structure of the solitons obtained from the
simulations of the partial differential equation (CCQGLE) is shown
in Figs. 1-4. In each of the solitons presented in Figs. 1-4 one of the parameters is perturbed while all the other remaining  parameters are chosen from the Table 1 (see the row for the snake soliton, which is the unperturbed case).
Thus, only the perturbed parameter is indicated in the captions to the figures while the other ones stay unchanged.
\begin{figure}[ht]\label{figure2}
\begin{center}
\begin{tabular}{ccc}
\resizebox*{0.25\textheight}{!}{\rotatebox{0}
{\includegraphics{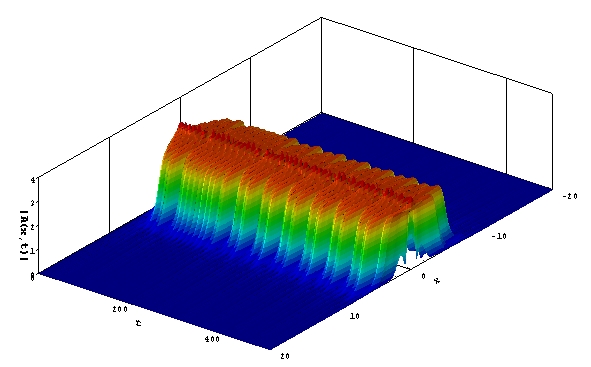}}}
&
\resizebox*{0.25\textheight}{!}{\rotatebox{0}
{\includegraphics{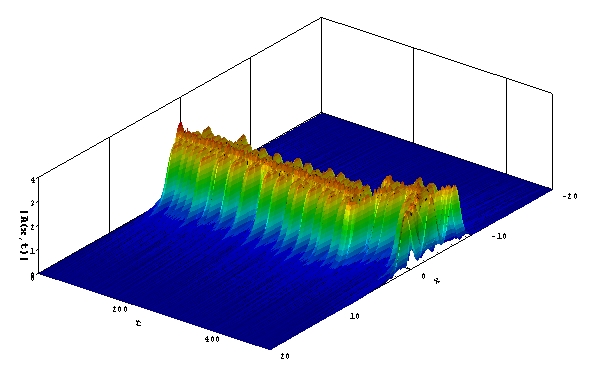}}}
\end{tabular}
\end{center}
\caption{Perturbed snaking soliton using the PDE: Left $b_3=-0.8$, Right $b_5=0.15$ }
\end{figure}
\begin{figure}[ht]\label{figure3}
\begin{center}
\begin{tabular}{ccc}
\resizebox*{0.25\textheight}{!}{\rotatebox{0}
{\includegraphics{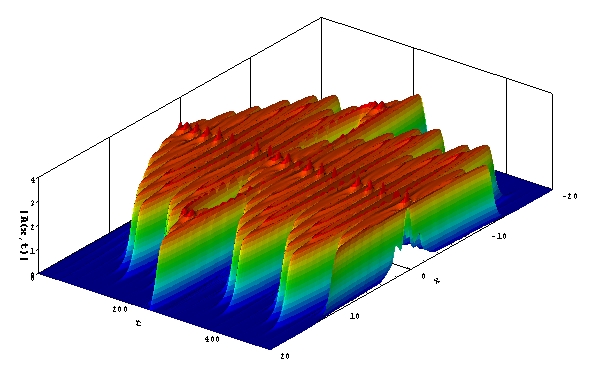}}}
&
\resizebox*{0.25\textheight}{!}{\rotatebox{0}
{\includegraphics{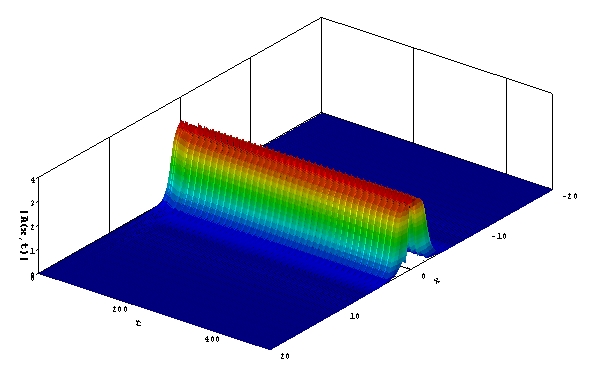}}}
\end{tabular}
\end{center}
\caption{Perturbed snaking soliton using the PDE: Left $c_1=0.55$, Right $c_3=1.135$}
\end{figure}
\begin{figure}[ht]\label{figure4}
\begin{center}
\begin{tabular}{ccc}
\resizebox*{0.25\textheight}{!}{\rotatebox{0}
{\includegraphics{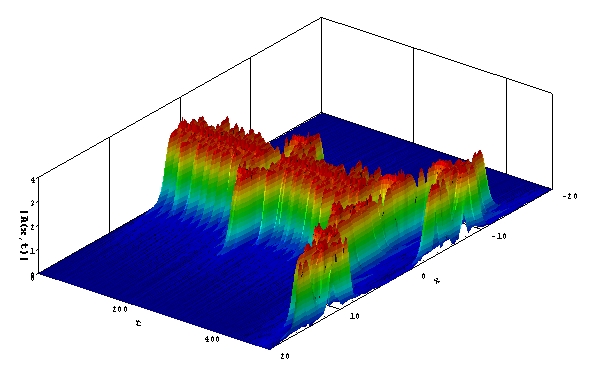}}}
&
\resizebox*{0.25\textheight}{!}{\rotatebox{0}
{\includegraphics{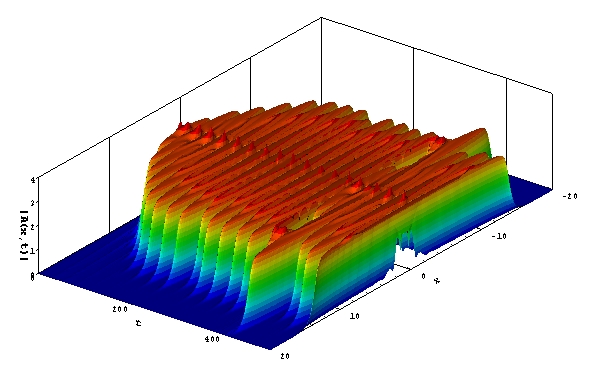}}}
\end{tabular}
\end{center}
\caption{Perturbed snaking soliton using the PDE: Left $c_5=-0.06$, Right $\epsilon=-0.08$}
\end{figure}
\section{Snake Solitons using Variational Approximation}
In our previous work \cite{Mancas:3}, we looked at stationary soliton solutions of the CCQGLE in the parameter regimes where they exist. However, these are not the only possible type of solutions. Pulsating solutions are another example of localized structures. They arise naturally from the stationary ones when the latter become unstable. Pulsating solitons might have several frequencies in their motions and the solutions will be quasiperiodic. A relative simpler case is when the motion has two frequencies.  Then, we will have a pulsating soliton \cite{Mancas:4}, which instead of having a zero velocity will move back and forward around a fixed point. Obviously, there are two frequencies involved in this motion which usually are incommensurate. We call this type of solitons with more than one frequency a snaking soliton. To capture this, we will use dynamical systems theory to construct solitons with quasiperiodic behavior.

Stable fixed points of the Euler-Lagrange system (\ref{4.5}) corresponds to the well-known
stationary solitons or plain pulses. The periodic evolution of $A_1(t)$, $\phi(t)$, $\alpha(t)$
on stable periodic attractors yields solitons whose amplitude is non-stationary.
To derive the conditions for occurrence of stable periodic orbits of $A_1(t)$, $\phi(t)$,
and $\alpha(t)$, we proceed as follows.

First, we fix a set of  system parameters
$b_1=0.08$, $b_5= 0.11$, $c_1=0.5$, $c_3=1$, $c_5=-0.08$ as in the case of the full PDE presented in the Table 1.
Then, we solve numerically the system of transcendental equations \eqref{4.5}, which are the equations of the fixed points. By the Ruth-Hurwitz conditions, the Hopf curve is defined as $\delta_1 \delta_2-\delta_3=0.$ This condition, along with the equations of the fixed points leads to onset of periodic solutions of \eqref{4.5} as we will see next.
\begin{figure}[ht]\label{Figure5}
\begin{center}
\begin{tabular}{cc}
\resizebox*{0.25\textheight}{!}{\rotatebox{0}
{\includegraphics{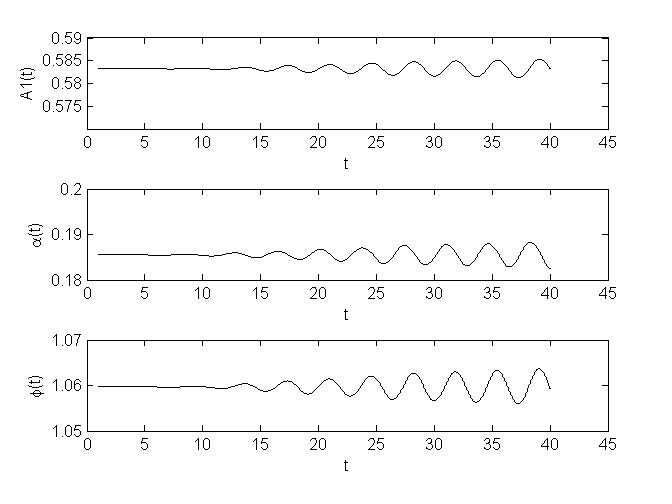}}}
&
\resizebox*{0.25\textheight}{!}{\rotatebox{0}
{\includegraphics{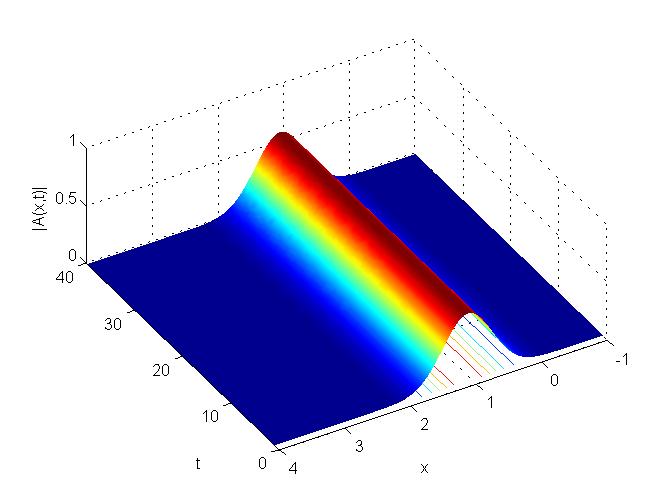}}}
\end{tabular}
\caption{Numerical simulation using the variational approximation: periodic time series of the amplitude $A_1(t)$,
the phase $\alpha(t)$ and the position $\phi(t)$ on the Hopf curve (Left), spatiotemporal structure of the amplitude
$|A(x,t)|=A_1^n e^{-\sigma^2\left(x_j-\phi^n\right)^2}$ on the Hopf curve (Right).}
\end{center}
\end{figure}
On the Hopf bifurcation curve we obtain that $b_{3Hopf}=-1.89646$, and $\epsilon_{Hopf}=-0.297393$, while the fixed points are $A_1(0)= 0.583236$, $\phi(0)=1.05969$, and $\alpha(0)=0.185515$. For these values of $b_{3Hopf}$ and $\epsilon_{Hopf}$, which are two free bifurcation parameters, we integrate numerically the systems of 3 ODEs \eqref{4.5}, using as initial conditions the three values of the fixed points, to find the trial function's periodic functions.  Hopf bifurcations occur in this system leading to periodic orbits. These are the conditions which exhibit  periodic oscillations, and hence stable snaking solitons will exist within our variational formulation.

Next, we plot the time series for the amplitude $A_1(t)$, the phase $\alpha(t)$ and the position $\phi(t)$ on the Hopf curve and the results are presented in Fig. 5 (Left). As expected, we noticed that the amplitude is small, since the maximum height of the snaking soliton is proportional to the square root of the distance from the Hopf curve. The corresponding spatio-temporal structure of the ansatz $|A(x,t)|=A_1^n e^{-\sigma^2\left(x_j-\phi^n\right)^2}$ on the Hopf curve is shown in Fig. 5 (Right). Notice in the figure very small undulations in amplitude.

To construct snake solitons with amplitudes large enough, we had to move away from the Hopf curve, as much as possible, but at the same time to be sure not to be outside of the parameters ranges for the existence of the snaking soliton. That could be achieved by varying one or more of the system parameters. First, we varied the first bifurcation parameter $\epsilon$, slowly away from the Hopf
curve where $b_3=b_{3Hopf}$, and $\epsilon=\epsilon_{Hopf}$, and we noticed that the snaking soliton had very small amplitude of $A_1(t)$. Since the soliton in this case had only a magnitude of only $10^{-4}$, we decided to vary the second bifurcation parameter $b_3$, which stands for the cubic gain when negative. We found that the domain of existence for the snaking soliton as a function of $b_3$ was $[-2.25234,-0.143456]$, passing through the Hopf curve value of $b_{3Hopf}=-1.89646$. For the unperturbed values of $b_3=-0.835$, and $\epsilon=-0.1$, we numerically integrate the Euler-Lagrange system of ODEs \eqref{4.5} using the Matlab's built-in function \verb+ode45+ and we plot the periodic orbit, which is shown in Fig. 6 (Left), while the resulting periodic time series for $A_1(t)$, $\alpha(t)$, and $\phi(t)$ are shown in Fig. 6 (Right).  The resulting time series are then substituted back in the ansatz \eqref{5.2} and the spatiotemporal structure $|A(x,t)|$ of the snaking soliton obtained by the variational formulation is presented in Fig. 7.
As the various system parameters $c_1$, $c_3$, $c_5$, $b_1$, $b_3$, $b_5$, and $\epsilon$ within the stable regime are varied, the effects of the amplitude, position, width (and, less importantly, phase) may also be studied, and this is discussed subsequently.
\begin{figure}[ht]\label{figure6}
\begin{center}
\begin{tabular}{ccc}
\resizebox*{0.25\textheight}{!}{\rotatebox{0}
{\includegraphics{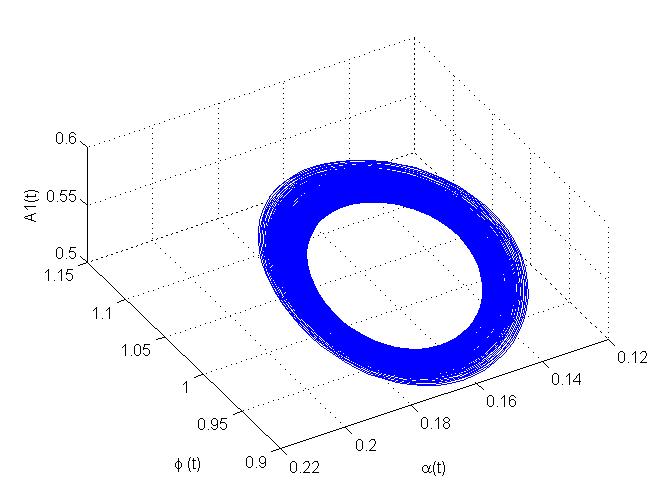}}}
&
\resizebox*{0.25\textheight}{!}{\rotatebox{0}
{\includegraphics{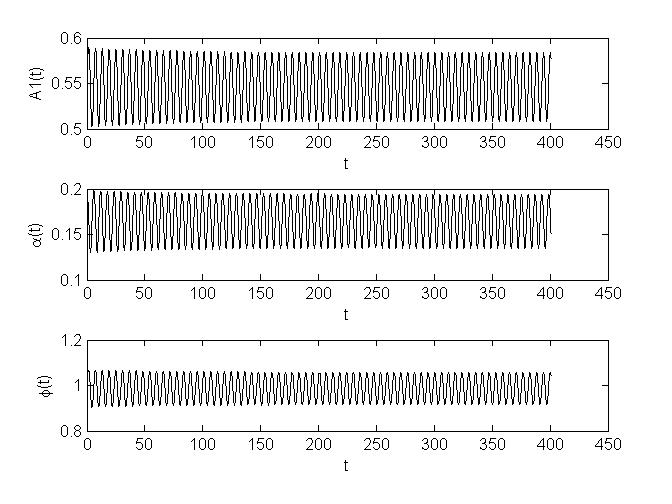}}}
\end{tabular}
\end{center}
\caption{Numerical simulation using the variational approximation for the snake of Fig. 1: periodic orbit or limit cycle (Left), periodic time series for $A_1(t)$, $\alpha(t)$, and $\phi(t)$ (Right)}
\end{figure}
\begin{figure}[ht]\label{figure7}
\begin{center}
\resizebox*{0.3\textheight}{!}{\rotatebox{0}
{\includegraphics{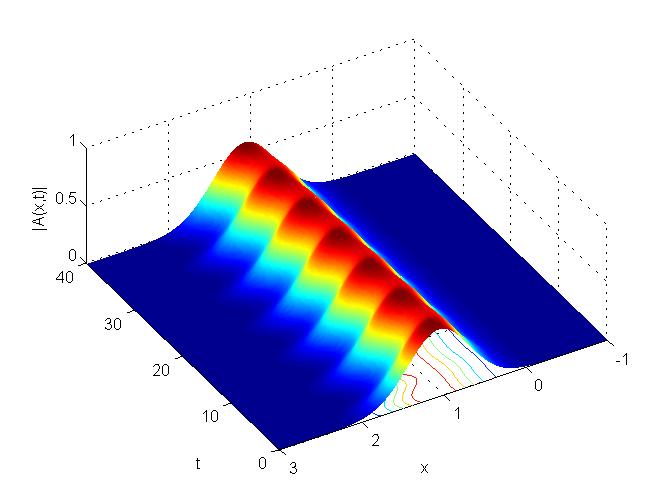}}}
\end{center}
\caption{Snake soliton using the ODEs (variational approximation): unperturbed case}
\end{figure}
As observed from the simulations of the full PDE and ODEs system, the snaking soliton obtained
from the ODE system (\ref{4.5}) qualitatively agree for a short time to the solution of the
complex nonlinear PDE (\ref{2.1}). Let us take a closer look to the snake solitons shown in Figs. 7--10.
The soliton  now ``snakes" or wiggles as its position varies periodically in time.
Note that the amplitude $|A(t)|$ varies periodically as $A_1(t)$ varies, but there
would be additional amplitude modulation due to the periodic variation of $\phi(t)$.
Next, we shall consider the effect of all the various parameters in the CCQGLE \eqref{2.1} on the shape (amplitude, position, phase, period) and stability of the snake.

In considering the parameter effects on snake shape and period, note that the wave is a spatially coherent structure (or a ``collective coordinate" given by the trial function) whose parameters oscillate in time. Hence, the temporal period of the snake is the same as the period $T$ of the oscillations of $A_1(t)$, $\phi(t)$, and $\alpha(t)$ on their limit cycle. As for the peak amplitude and peak position of the snake, these are determined by the peak amplitude $A_{1p}$ of $A_1(t)$, and the peak position $\phi_p$ of $\phi(t)$ respectively. Notice that from \eqref{5.2} we can regard the width and the amplitude of the snake as being inversely proportional to position $\phi(t)$ for the snake i.e., at any time $t$ when the amplitude is minimum, the width will be minimum, so the position is maximum and vice versa. So, maximum deflection from the horizontal position $x=const.$ is obtained when the position of the snake is maximum, and hence the width and amplitude are minimum.

Keeping the above in mind, we vary the parameters of the CCQGLE in turn and we observe the resulting effects on $A_{1p}$ (the peak amplitude), $\phi_p$ (the position), and $T$ (the temporal period) of the snake soliton:
\begin{enumerate}
\item[(i)] For \textit{increased} $b_1$, the values of $A_{1p}$, $\phi_p$, and $T$ all \textit{increase}.\\
\item[(ii)] \textit{Increasing} $b_3$ \textit{augments} all of $A_{1p}$, $\phi_p$, and $T$.\\
\item[(iii)] \textit{Increasing} $b_5$ \textit{increases} all of $A_{1p}$, $\phi_p$, and $T$.\\
\item[(iv)] \textit{Raising} $c_1$ \textit{increases } $A_{1p}$, $\phi_p$, but \textit{decreases} $T$.\\
\item[(v)] \textit{Incrementing} $c_3$ \textit{decreases} all of $A_{1p}$, $\phi_p$, and $T$.\\
\item[(vi)] \textit{Augmenting} $c_5$ causes a \textit{decrease} in $A_{1p}$, $\phi_p$, and \textit{increases} $T$.\\
\item[(vii)] \textit{Raising} $\epsilon$ causes  $A_{1p}$, $\sigma_p$ to \textit{rise}, but $T$ to \textit{fall}.
\end{enumerate}
\begin{figure}[ht]\label{figure8}
\begin{center}
\begin{tabular}{ccc}
\resizebox*{0.25\textheight}{!}{\rotatebox{0}
{\includegraphics{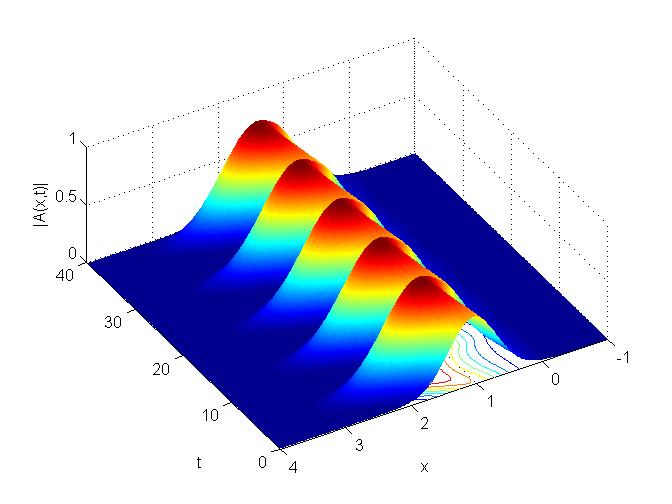}}}
&
\resizebox*{0.25\textheight}{!}{\rotatebox{0}
{\includegraphics{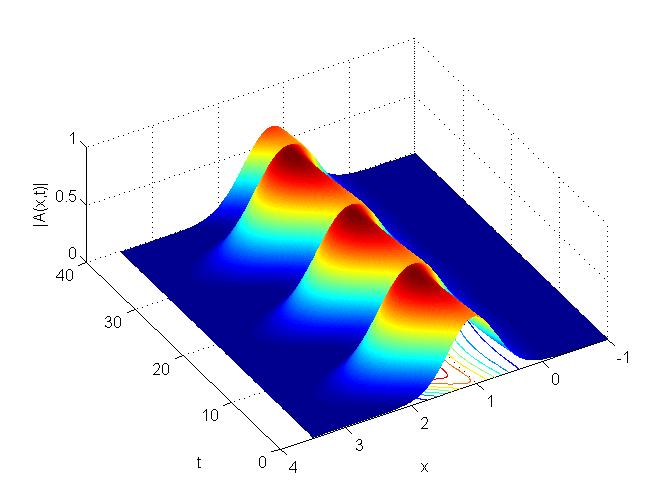}}}
\end{tabular}
\end{center}
\caption{Snake soliton using the ODEs (variational approximation): Left $b_3=-0.6$, Right $b_5=0.62$}
\end{figure}

\begin{figure}[ht]\label{figure9}
\begin{center}
\begin{tabular}{ccc}
\resizebox*{0.25\textheight}{!}{\rotatebox{0}
{\includegraphics{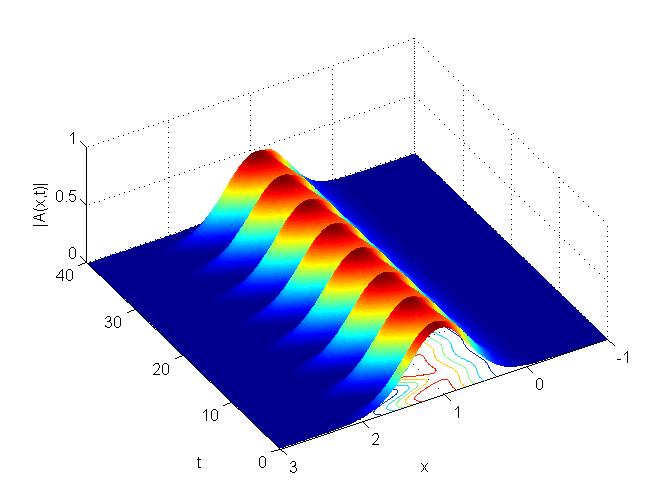}}}
&
\resizebox*{0.25\textheight}{!}{\rotatebox{0}
{\includegraphics{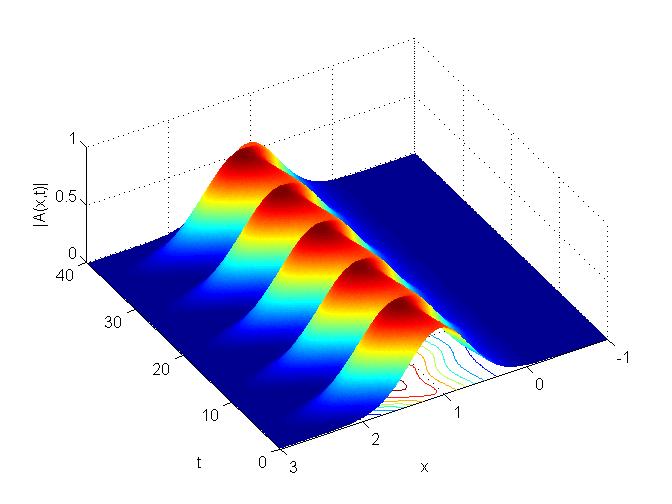}}}
\end{tabular}
\end{center}
\caption{Snake soliton using the ODEs (variational approximation): Left $c_1=0.55$, Right $c_3=2$}
\end{figure}

\begin{figure}[ht]\label{figure10}
\begin{center}
\begin{tabular}{ccc}
\resizebox*{0.25\textheight}{!}{\rotatebox{0}
{\includegraphics{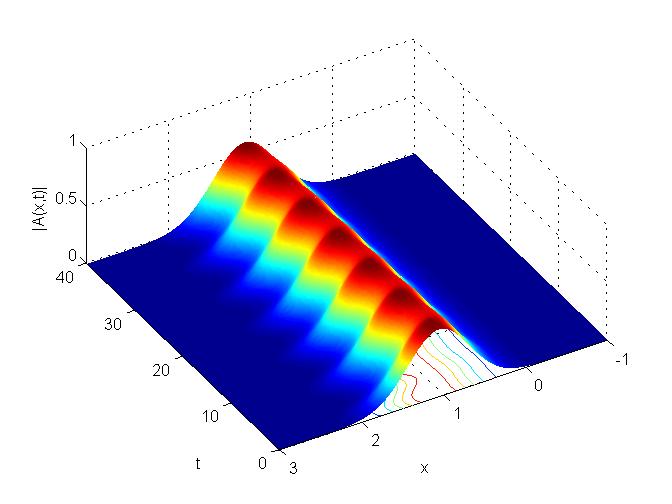}}}
&
\resizebox*{0.25\textheight}{!}{\rotatebox{0}
{\includegraphics{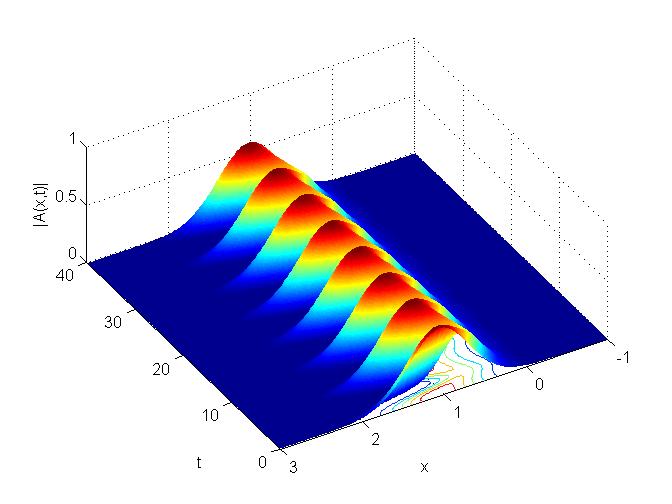}}}
\end{tabular}
\end{center}
\caption{Snake soliton using the ODEs (variational approximation): Left $c_5=0.8$, Right $\epsilon=-0.08$}
\end{figure}

The above constitute our detailed predictions of the various parameters in the CCQGLE on the amplitude, position, and temporal width of the snake solitons. We have verified that each set of predictions (i)-(vii) above agree when the corresponding parameter is varied in the solitary wave simulation for the full PDE. Note also that $A_1(t)$ and $\phi(t)$ are always in phase, so that $A_{1p}$ and $\phi_p$ occur simultaneously. Thus, the snaking solitons are tallest where they have the greatest width. This is completely consistent with our simulation, as well as those in \cite{Artigas}.
\section{Conclusion}\label{S:5}
In this paper we discussed a brief theoretical framework for
analyzing the full spatiotemporal structure of one class of solitary
waves (snaking soliton) in the CCQGLE and presented some numerical simulations of
dissipative snaking soliton solutions to the CCQGLE.
The results obtained analytically using the
variational approximation for the snaking soliton is compared with
the numerical simulations of the CCQGLE.
The specific theoretical modeling includes the use of a recent variational
formulation and significantly generalized trial function for the solitary waves
solutions. In addition, the resulting Euler-Lagrange equations are
treated in an entirely different way by looking at their stable
periodic solutions (or limit cycles) resulting from supercritical
Hopf bifurcations. Oscillations of their trial function parameters
on these limit cycles provide the pulsations of the amplitude,
width/position, and phase of the solitons. The model also allows for
detailed predictions regarding the other types of solitons i.e.,
pulsating, chaotical and creeping.

In the future, we will explore on the (2D) solutions called
spinning solitons as well as the  (3D) solutions called
optical bullets. These depict a confined spatiotemporal soliton in
which the balance between the focusing nonlinearity and the spreading
while propagating through medium provides the shape of a bullet
\cite{Crasovan,Soto:3}.
\bibliographystyle{plain}
\bibliography{bib9}
\end{document}